\documentclass[useAMS]{mn2e}

\usepackage{graphicx}
\usepackage{times}

\begin{document}


\newcommand{\intdv}{$\int$\trstar$\d v$}
\newcommand{\hcop}{HCO$^{+}$}
\newcommand{\htcop}{H$^{13}$CO$^{+}$}
\newcommand{\htcn}{H$^{13}$CN}
\newcommand{\kms}{$\,$km$\,$s$^{-1}$}
\newcommand{\mic}{$\mu$m}
\newcommand{\cucm}{cm$^{-3}$}
\newcommand{\sqcm}{cm$^{-2}$}
\newcommand{\degs}{$^{\circ}$}
\newcommand{\msun}{M$_{\odot}$}
\newcommand{\lsun}{L$_{\odot}$}
\newcommand{\tastar}{$T_{\rm A}^{*}$}
\newcommand{\trstar}{$T_{\rm R}^{*}$}
\newcommand{\trad}{$T_{\rm R}$}
\newcommand{\amm}{NH$_{3}$}
\newcommand{\too}{$\rightarrow$}
\newcommand{\ceo}{C$^{18}$O}
\newcommand{\cso}{C$^{17}$O}
\newcommand{\thco}{$^{13}$CO}
\newcommand{\twco}{$^{12}$CO}
\newcommand{\ci}{C\,{\textsc i}}

\title[Collimated outflows in G35.2--0.7N]{A detailed study of
  G35.2--0.7N: collimated outflows in a cluster of high-mass young
  stellar objects}

\author[A.G. Gibb et al.] 
{A.G. Gibb$^{1,2}$\thanks{Email: agg@astro.umd.edu}, M.G. Hoare$^2$,
L.T. Little$^{3}$ and M.C.H. Wright$^{4}$\\ 
$^1$Department of Astronomy, University of Maryland, College
Park, MD 20742, USA. \\
$^2$Department of Physics and Astronomy, University of Leeds, Leeds,
West Yorkshire, LS2 9JT, UK. \\
$^3$Electronic Engineering Laboratory, University of Kent at
Canterbury, Kent, CT2 7NT, UK.\\
$^4$Radio Astronomy Laboratory, Department of Astronomy, University of
California, 601 Campbell Hall, Berkeley, CA 94720, USA. 
}

\maketitle

\begin{abstract}
  We present a series of JCMT, BIMA and VLA observations of the
  massive star-forming region associated with G35.2--0.7N. These new
  observations shed considerable light on the nature of the outflows
  in this region. The combination of our CO, SiO and radio data
  suggest that there are perhaps as many as four outflows emanating
  from the core containing G35.2--0.7N.
  
  CO $J$=3\too 2 maps show that the outflow has a curved appearance
  consistent with precession of a central driving source. However, the
  geometric centre of the flow is found to be not coincident with the
  radio jet source G35.2N but is instead closer to a peak in SiO,
  H$^{13}$CO$^+$ and dust continuum found in the BIMA data. An
  elongated finger of CO emission is detected to the north of centre
  which points back towards the radio jet centred on G35.2N, further
  ruling it out as the driving source of the larger-scale CO flow.
  
  BIMA observations of the 3.5-mm continuum (which is dominated by
  dust), \htcn\ and H$^{13}$CO$^+$ emission trace a dense, elongated,
  rotating envelope with properties in good agreement with values
  derived from previous ammonia observations. The peak of the dust
  continuum and the H$^{13}$CO$^+$ peak a few arcsec to the south of
  G35.2N. SiO $J$=2\too 1 data delineate a well-collimated feature
  parallel with the axis of the CO outflow, but offset to the north by
  $\sim$10 arcsec. \htcop\ emission is detected at the possible origin
  of this flow but no radio source is observed.
  
  VLA A-configuration observations at 6- and 3.5 cm resolve the radio
  jet into at least six discrete components, with positions consistent
  with previous observations. The central driving source, G35.2N, is
  only detected at 3.5 cm. At least two other sources are detected,
  one of which lies within the flattened core and may be associated
  with another flow inferred from recent $L'$-band observations. No
  radio source is detected at the geometric centre of the CO outflow.
\end{abstract}


\section{Introduction}

Outflows from low mass young stellar objects (YSOs) have been
extensively studied, especially with the advent of sensitive,
high-resolution millimetre facilities. A number of spectacular CO
outflows have been mapped (e.g. Gueth \& Guilloteau 1999, Lee et al.
2000) and the current consensus is that these outflows are driven by
highly collimated jets, which are sometimes seen optically or in the
emission of shock-excited H$_2$ or shock-ionized gas via radio
continuum (Eisl\"offel 1997; Rodr\'{\i}guez 1997). The CO traces the
gas swept up by the outflow as it ploughs into the surrounding cloud
and effectively provides a time-integrated historical view of the flow
evolution. The H$_2$ emission arises from within the shocked regions
of the jet and provides an instantaneous view of the current
interaction regions, as its cooling time is of order a year or so
(Hollenbach \& McKee 1979).

The study of outflows from high-mass YSOs is more problematic than for
low mass sources for several reasons. Massive stars form in dense
clusters and even the nearest massive YSOs lie at greater distances
than typical low-mass star forming cores, both of which mean that
confusion makes it difficult to correctly interpret observations of
massive star forming regions.  The contraction timescales of high-mass
stars are short and their smaller numbers further renders the study of
outflows from luminous YSOs challenging. For example, it is not clear
whether high mass outflows are driven by collimated jets as evidence
for their existence is scant. At present there appears to be only a
few examples of luminous YSOs with well-defined jets such as HH80/81
(Mart\'{\i} et al. 1993), Cep A2 (Torrelles et al. 1996) and W3-H$_2$O
(Wilner, Reid \& Menten 1999). A recent survey of outflows form
massive YSOs by Beuther et al. (2002) found a higher degree of
collimation than previously observed. However, the majority of massive
YSOs do not show evidence for jets. 

Surprisingly some massive YSOs (notably S106 and S140-IRS1) actually
appear to have winds which extend in the equatorial direction (Hoare
2002; Hoare et al. 1994).  The radio emission from massive YSOs of
ZAMS type B0 and later tends to be dominated by free-free emission
from an ionized stellar wind rather than an H\,{\sc ii} region.
Free-free emission from a stellar wind has a spectral index of +0.6
(Wright \& Barlow 1974), compared with +2 (--0.1) for an optically
thick (thin) H\,{\sc ii} region. The spectral index of a stellar wind
may steepen to $\sim$1 or higher if the wind recombines at a finite
distance from the star (Simon et al.  1983). An accelerated wind may
also result in a spectral index greater than 0.6 (Panagia \& Felli
1975).

The target for this present study is G35.2--0.7N, a massive
star-forming region in which a jet-like radio structure is seen.
G35.2--0.7N is associated with the IRAS source 18556+0136 (and
hereafter will be referred to as G35.2N) lies approximately 2 kpc
distant. Dent et al. (1985) derive a spectral class of B0.5 for the
exciting star of the bipolar reflection nebula, corresponding to a
luminosity of approximately 10$^4$\, L$_\odot$ and a stellar mass of
$\sim$15 M$_\odot$ (Allen 1973). A rotating interstellar-scale disc or
toroid was detected in ammonia emission (Little et al. 1985) while a
bipolar CO outflow was discovered to lie almost perpendicular to this
disc (Dent et al.  1985). Heaton \& Little (1988, hereafter HL88) made
VLA observations of the radio emission at the centre of the outflow
and revealed an elongated source orientated north-south, a position
angle very different from that of the CO flow.  HL88 resolved three
main radio components arranged in a co-linear fashion labelled N, C
and S. The centre component (C) is coincident with a cluster of OH
masers (Brebner et al. 1987) and has a spectral index of 0.8,
indicating a stellar wind (Wright \& Barlow 1974), while the others
have spectral indices close to zero, suggesting optically thin H{\sc
  ii} regions.  The symmetry of the radio emission suggests a jet but
the spectral indices are consistent with a number of embedded YSOs.

Also orientated north-south in G35.2N is a bipolar reflection nebula
(the brightest part of which had previously been called G35.2N-IRS1),
studied in detail by Walther, Aspin \& McLean (1990). These authors
speculated that a source G35N-star located in a region of high
extinction was responsible for the nebula. The location of this
proposed source is approximately 3 arcsec south of the radio source
G35.2N. More recent $L'$-band infrared observations by Fuller,
Zijlstra \& Williams (2001) showed that the reflection nebula is still
visible at $L'$-band, although its appearance supports the idea of a
north-south jet emerging from G35.2N.

HL88 proposed that the large difference in position angle between the
CO and radio was due to precession of a collimated, ionized jet, an
interpretation given further support by Little, Kelly \& Murphy
(1998). These authors observed that high-velocity neutral atomic
carbon was confined to a region with an intermediate position angle,
indicating perhaps that ultraviolet radiation from the central star
has penetrated into a cavity swept clear by the outflow (seen as the
infrared reflection nebula), dissociating CO molecules on the near
side of the shell. G35.2N is not alone in showing evidence for
precession. The jet in HH80--81 is probably precessing, albeit through
a much smaller angle (Mart\'{\i} et al. 1993), while Shepherd et
al. (2000) invoke precession to explain the large angular offset
between the large- and small-scale outflows in IRAS\,20126+4104.

To explore further the precession hypothesis for G35.2N and determine
the nature of the radio sources, we have carried out a detailed
observational study of this outflow source in which we have mapped the
outflow and disc in a variety of molecular tracers with the JCMT and
BIMA as well as carried out high-resolution VLA observations of the
jet itself. In the following sections we describe our observations and
ultimately arrive at a new picture for the region containing G35.2N.

\section{Observations}

\begin{figure*}
\centering 
\includegraphics[width=15.5cm]{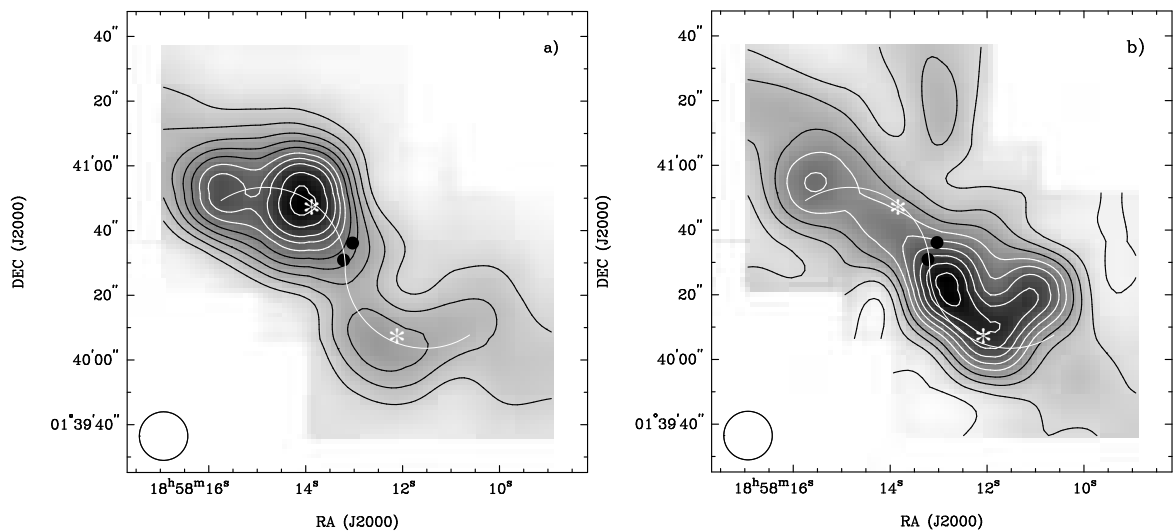}
\caption{CO $J$=3\too 2 emission integrated between a) 5 and 30 \kms\
  and b) 40 and 55 \kms. The redshifted emission is only half as
  bright as the blueshifted lobe (partly due to its smaller velocity
  coverage).  The contours start at 25 (20) K~\kms, increasing in
  steps of 25 (10) K~\kms for the blueshifted (redshifted) emission.
  The greyscale is linear between 0 and 250 K\,km\,s$^{-1}$ (blue) and
  0 and 110 K\,km\,s$^{-1}$ (red). The radio source G35.2N at (0,0)
  and millimetre source G35MM2 at (2.8,--5.2) arcsec offset are each
  marked with filled circle. The position of G35MM2 is taken from
  J.M. Carpenter (private communication).
  The white curve marks an `S' shape which might be expected for a
  precessing source. The asterisks mark the positions `B' and `R'
  referred to in the spectra displayed in Fig.~\ref{fig:spectra}. The
  large open circle represents the 15-arcsec JCMT beam.
\label{fig:co32int}}
\end{figure*}

Observations were made at the 15-m James Clerk Maxwell Telescope
(JCMT), Mauna Kea, Hawaii, the Berkeley-Illinois-Maryland Association
(BIMA) array of nine (now ten) 6.1-m telescopes at Hat Creek,
California, and the NRAO Very Large Array (VLA), New Mexico. From
previous observations (e.g. Little et al. 1985) the systemic velocity
is +34 \kms\ with respect to the local standard of rest (LSR).

The JCMT observations were carried out on the nights of 1995 April 16
(\cso\ only) and 1997 April 17 to 20 and made use of the common user
receivers A2 (covering 200 to 280 GHz), B3i (300 to 380 GHz), B3 (315
to 370 GHz) and C2 (450 to 500 GHz). In the 1995 run, the weather was
excellent with 225-GHz optical depths of $<$0.07. G34.3+0.2 and
AFGL2591 were used as pointing sources (good to within 2 arcsec), and
G34.3+0.2 was used as a secondary calibrator on which we obtained
standard spectra. In this run, receiver B3i was used (double sideband)
which had a system temperature of around 750 K, with the digital
autocorrelation spectrometer (DAS) set to a bandwidth of 500 MHz
(giving a spectral resolution of 378 kHz). The calibration uncertainty
is estimated at 10 per cent. The (0,0) position in the JCMT maps is at
RA(J2000) = 18$^{\rm h}$58$^{\rm m}$13.00$^{\rm s}$, Dec(J2000) =
+01$^\circ$40$'$36.2$''$

In the 1997 run the weather conditions were reasonably good (the
optical depth at 225 GHz was typically 0.12) with pointing errors at
$\sim$2 arcsec or less. Mars and G34.3+0.2 were used as pointing
sources, with G34.3+0.2 used as a secondary calibrator. The
instrumental setup employed the facility receivers A2, B3 (in both
single- and double-sideband operation) and C2 in conjunction with the
DAS as backend in the same configuration as above. Typical system
temperatures were 380 K (A2), 600--1500 K (B3) and 3500 K (C2).
Integration times were typically 5 to 10 minutes per point depending
on system temperature and receiver. The calibration uncertainty is
estimated at 10--20 per cent, again depending on receiver (with the
smaller uncertainty corresponding to lower frequency data). Linear
baselines were removed from all JCMT spectra. The target species were
CO, \thco\ and \cso\ $J$=3\too 2 (Rx B3) and SiO $J$=5\too 4 (A2).
The full-width-at-half-maximum (FWHM) diameter of the JCMT beam is 22
arcsec at 217 GHz (for the SiO 5\too 4 line), 15 arcsec at 345 GHz
(3\too 2 CO, $^{13}$CO and C$^{17}$O) and 10 arcsec at 461 GHz (CO
4\too 3). All JCMT brightness temperatures are quoted in the text as
\trstar\ assuming a value for the forward scattering and spillover
efficiency ($\eta_{\rm fss}$) of 0.8, 0.7 and 0.7 at each of the three
frequencies given above.

The BIMA observations were carried out in a single track (with 5 hours
on source) on 1997 April 18 with the array in its C configuration. The
flux calibrator was 3C273 (assuming a flux density at 86 GHz of 32
Jy). The phase calibrator was 1751+096 (J2000) and was observed every
30 minutes. A flux density of 4.3 Jy was derived for 1751+096,
bootstrapped from the flux density of 3C273. The instrumental setup
permitted simultaneous observation of the SiO $J$=2\too 1,
H$^{13}$CO$^+$ $J$=1\too 0 and the hyperfine triplet of H$^{13}$CN
$J$=1\too 0, $F$=1\too 1, 2\too 1 and 1\too 0 lines with a bandwidth
and spectral resolution of 25 MHz and 98 kHz (0.34 \kms) respectively.
In addition a continuum band of 100 MHz was observed which was used to
construct the 3.5-mm continuum image.  Line-free channels were
averaged and subtracted from the line data before images were made.
Data calibration and reduction was performed using standard procedures
within {\sc miriad}. The size of the restoring {\sc clean} beam was
10$\times$8 arcsec$^2$ (uniform weighting), rising to 11$\times$9
arcsec$^2$ for natural weighting.  Most images were formed with
natural weighting to maximize
sensitivity. The pointing centre for the BIMA observations is at
RA(J2000) = 18$^{\rm h}$58$^{\rm m}$13.13$^{\rm s}$, Dec(J2000) =
+01$^\circ$40$'$39.50$''$.

The VLA observations were carried out on 1999 September 15 in the A
configuration. G35.2N was observed at C, X and U bands with a total
bandwidth of 200 MHz.  The primary flux calibrators were 3C286 (C:
7.47 Jy) and 3C48 (X and U: 3.25 and 1.78 Jy respectively) and the
phase calibrator was 1849+005 (J2000). The bootstrapped flux densities
of the secondary calibrator were 0.83 Jy (C), 0.85 Jy (X) and 0.92 Jy
(U). After editing and calibration the data were imaged using the {\sc
  aips} task {\sc imagr} with uniform weighting to maximize the
resolution. The 5-GHz data were spatially filtered to only include
data on baselines longer than 50 k$\lambda$ to avoid contamination by
the bright H\,{\sc ii} region G35.2S (see HL88). The half-power
dimensions of the {\sc clean} beams were 0.57$\times$0.37 arcsec$^2$
at a position angle of 49 degrees (C), 0.25$\times$0.22 arcsec$^2$ at
33 degrees (X) and 0.14$\times$0.13 arcsec$^2$ at 35 degrees (U). The
VLA pointing centre is at RA(J2000)=18$^{\rm h}$58$^{\rm
  m}$12.93$^{\rm s}$, Dec(J2000)=+01$^\circ$40$'$36.5$''$.
Unfortunately high decorrelation in the U-band data meant that no
images could be made.

\begin{figure}
\centering
\includegraphics[width=8cm]{g35_fig02.eps}
\caption{ Example spectra from three positions in
  Fig.~\ref{fig:co32int}. Left panel is from peak position in the blue
  lobe (marked by an asterisk), centre panel is from the position of
  G35.2N and the right panel is from position marked by an asterisk in
  the red (south west) lobe. From
  top to bottom in each panel: \twco\ 4\too 3, 3\too 2 and \thco\ 
  3\too 2. The $x$-axis scale is velocity with respect to the LSR; the
  systemic velocity is +34 \kms. The dip in the $^{12}$CO spectra at
  15 \kms\ arises from a foreground cloud.  The `bumps' in the \thco\ 
  spectrum at the centre position are lines of methanol observed in
  the upper sideband. Note the different vertical scale for the \thco\
  spectra. The brightness temperatures in the figure may be converted
  to \trstar\ assuming $\eta_{\rm fss}$=0.7. } 
\label{fig:spectra}
\end{figure}

\section{Results}

\subsection{The CO outflow}

\begin{figure*}
\centering 
\includegraphics[width=15cm]{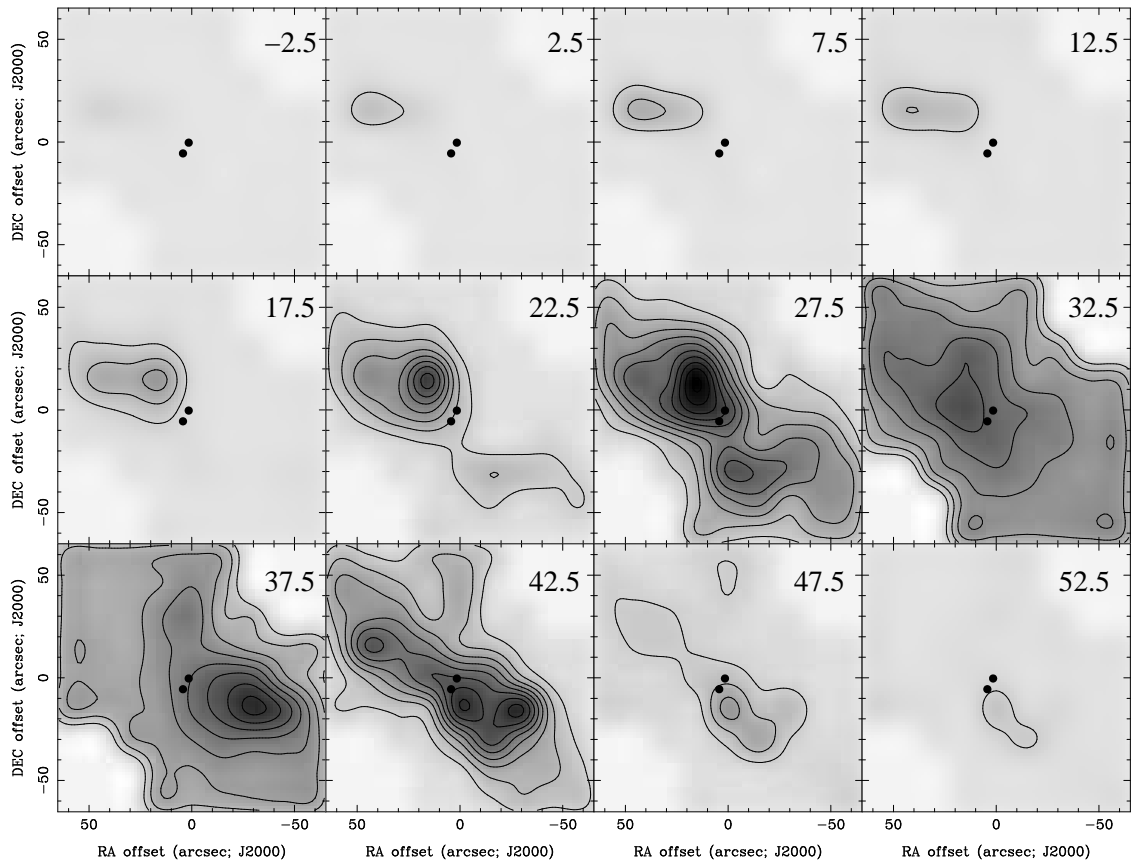}
\caption{CO $J$=3\too 2 emission integrated over successive
  5-\kms-wide velocity channels between 5 and 50 \kms. Lowest contour
  and interval are 10 and 10 K~\kms. The centre velocity of each panel
  is given in the top right corner. The two filled circles mark the
  positions of G35.2N at (0,0) and G35MM2 at (2.8,--5.2) arcsec
  offset.
\label{fig:co32chan}}
\end{figure*}

Previous interpretation of the CO data has assumed the presence of a
single driving source. The combination of our observations and other
results show that there are multiple sources in this region, and it is
likely that all contribute towards the observed CO distribution. The
3\too 2 CO blue- and red-shifted integrated intensity maps
(Fig.~\ref{fig:co32int}) have a similar appearance to the 1\too 0 maps
of Dent et al. (1985). Since the 3\too 2 line traces warmer and denser
material than the 1\too 0 the detailed outflow structure is delineated
more precisely. From these observations we find that the geometric
centre of the outflow does not appear to coincide with the radio
source G35.2N, and instead lies a few arcsec to the south-east. A new
millimetre continuum source (see below) is a better candidate. Example
CO 3\too 2 and 4\too 3 and \thco\ 3\too 2 spectra observed towards the
blue and red lobes and central position are shown in Fig.~\ref{fig:spectra}.

The outflow clearly lies close to the plane of the sky, although the
fact that the blueshifted emission in the north-east lobe is stronger
than that in the south-west lobe (and vice versa for the redshifted
emission) shows that this lobe is inclined slightly towards us. The
near-infrared reflection nebula (Walther et al. 1990) supports this
geometry with its greater extent to the north of G35.2N.  The apparent
opening angle of the CO flow is $\sim$45 degrees.  Assuming a
biconical geometry the inclination probably lies between 0 and 23
degrees relative to the plane of the sky. Since there is substantial
overlap between the red- and blue-shifted emission, an angle closer to
0 than 23 degrees seems most likely.

The blue lobe shows an S-like shape which is similar to what would be
expected for an outflow driven by a precessing source (e.g. Cliffe et
al. 1995). The red lobe does not have the same shape and in fact
appears to be more collimated than the blue lobe with an unresolved
jet-like appearance. A new feature not seen in the 1\too 0 data is the
elongated red-shifted component extending to the north of centre. Due
to confusion, it is difficult to tell whether it has a blue-shifted
equivalent. This flow may be the molecular counterpart to the
collimated radio jet (see below), and reinforces the interpretation
that there are a number of separate outflows emanating from within the
same molecular envelope, albeit with different orientations. Clearly
higher-resolution observations are necessary to disentangle the
outflow geometry near the driving sources.

Channel maps of the 3\too 2 emission are presented in
Fig.~\ref{fig:co32chan}. The lobes are not quite symmetric in their
appearance with the north-east lobe extending beyond the edge of the
mapped area. Furthermore the blue-shifted emission extends to a
greater velocity relative to the LSR velocity than the red-shifted
gas, 34 \kms\ (blue) compared with 15 \kms\ (red). The total velocity
extent of the outflow to a noise level of 0.2 K is $\sim$50 \kms. The
emission at the highest velocity is confined to a couple of `hot
spots' in each lobe of the flow, although they are not located
symmetrically about the central source. The low-velocity material
(best seen in the 32.5 \kms\ panel in Fig.~\ref{fig:co32chan}) appears
to delineate the edge of a cavity swept out by the outflow(s).

A number of CO 4\too 3 spectra were recorded in order to facilitate
calculation of the outflow parameters. Although severely self-absorbed
(see Fig.~\ref{fig:spectra}), the 4\too 3 spectra indicate the
presence of very warm gas. The shoulders of the self-absorbed 4\too 3
profiles imply temperatures of 40 K at the centre and in excess of 20
K within the outflow lobes (assuming the emission uniformly fills the
beam). The 4\too 3 spectra also show the greatest velocity extent of
all the CO spectra with emission extending to $-$45 \kms\ relative to
the line centre at the eastern CO peak and +28 \kms\ in the south-west
lobe.

\subsection{CO isotopomers}

\begin{figure}
\centering 
\includegraphics[width=8cm]{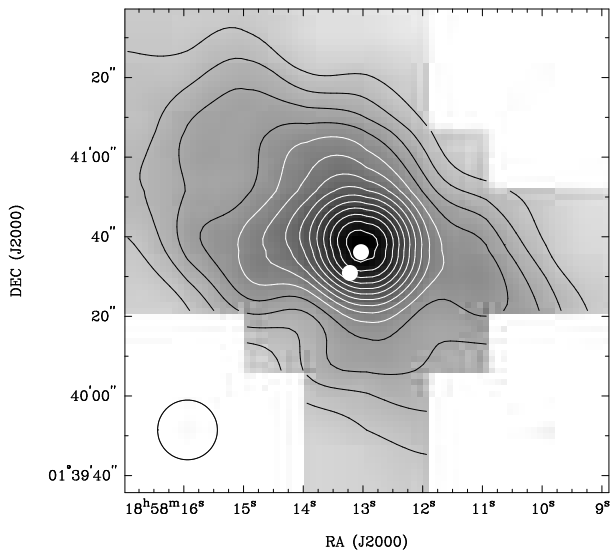}
\caption{\thco\ $J$=3\too 2 emission integrated between 25 and 45
  \kms. The contours start at 20 K~\kms\ increasing in steps of 5
  K~\kms. The radio source G35.2N and millimetre source G35MM2 are
  each marked with a filled circle as in Fig.~\ref{fig:co32int}. The
  open circle in the bottom left hand corner represents the JCMT beam.
\label{fig:13co32int}}
\end{figure}

The \thco\ 3\too 2 data (Fig.~\ref{fig:13co32int}) are dominated by a
central peak of emission coincident with the radio source G35.2N. The
line is strongly self-absorbed at the centre, and many of the spectra
show signs of self-absorption across the region mapped. The peak
antenna temperature (\trstar) is 14 K, indicating a minimum excitation
temperature of 21 K, although the line is heavily self-absorbed.
Spectra are shown in Fig.~\ref{fig:spectra}.  At low velocities the
\thco\ also appears to trace the edge of the outflow cavity. From
these data, the cavity has the appearance of a hollow cylinder,
suggesting that the outflow does not continue expanding radially and
that at least some of the collimation of the flow may be effected on
large scales by the surrounding environment.  At higher velocities,
the central core of the flow is seen. The total velocity coverage to a
noise level of 0.2 K is 18 \kms.

Emission from the \cso\ 3\too 2 line is confined to a single symmetric
spectrum at the central position with a brightness temperature of 0.67
K and linewidth 6.2 \kms. The noise level in these data is 0.1 K per 1
MHz channel (0.9 \kms\ at 337 GHz).

\begin{figure}
\centering 
\includegraphics[width=8cm]{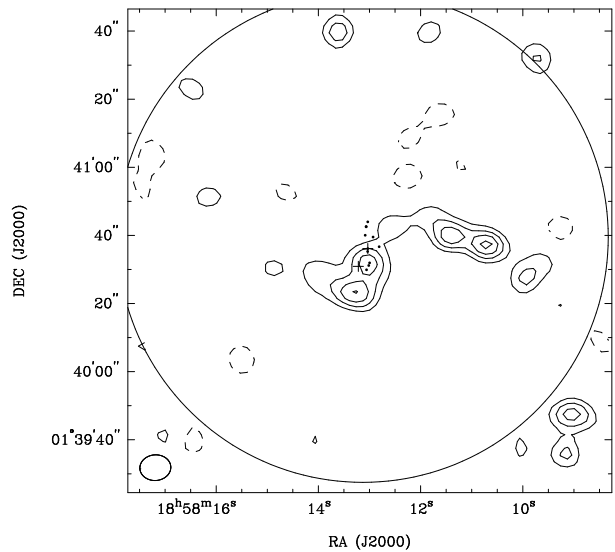}
\caption{SiO $J$=2\too 1 emission integrated over the velocity range
  31 to 37 \kms. The contours are --3$\sigma$, 3$\sigma$, then every
  2$\sigma$ to 15$\sigma$ ($\sigma$ = 0.8 K\,\kms). The crosses mark
  the radio source G35.2N and millimetre source G35MM2.  Small filled
  circles mark the position of the radio sources from
  Table~\ref{tab:radio1}. The large circle is the primary beam FWHM of
  the BIMA antennas at 86 GHz and the ellipse represents the {\sc
    clean} beam.
\label{fig:sio21}}
\end{figure}

\subsection{SiO 2\too 1 and 5\too 4 emission}

The SiO 2\too 1 emission is shown in Fig.~\ref{fig:sio21} integrated
between 32 and 36 \kms. A number of interesting features are evident.
Rather than show up in the outflow, the peak SiO emission appears
located within the envelope around G35.2N, although it is offset by
several arcsec to the south-east from the radio source, closer to the
millimetre source G35MM2 (see below).  There is no SiO emission which
can be directly associated with the radio jet seen by HL88 (also see
section 3.4 below). The SiO peak at (4,--13) arcsec offset from G35.2N
does not appear to be associated with any known feature, although it
does lie in a line which extends beyond the southern group of radio
components. Further north towards the periphery of the BIMA primary
beam half-power point, a knot of SiO emission lies along a line which
can be traced back to the radio jet, some 50 arcsec to the south, and
appears to be associated with the red-shifted spur seen in the CO
data. This SiO maximum is also red-shifted, lying at $\sim$36\,\kms.

In addition, the 2\too 1 SiO shows a linear feature composed of
several knots of emission, extending parallel to the main CO outflow.
This SiO feature may be tracing the shocked edge of a swept-out
cavity, or it may be a separate well-collimated flow (see
Fig.~\ref{fig:sio21}). Although not unprecedented, there is no
evidence for the opposite lobe in this proposed outflow, neither in
our CO nor in SiO. Higher-resolution CO observations may help
distinguish it from the main CO flow.

Further observations are necessary to determine the reality of the
components far from the phase centre, in particular the northern SiO
knot and the linear feature. The majority of the SiO 2\too 1 emission
seen in Fig.~\ref{fig:sio21} is blue-shifted. The exception is the
emission associated with the envelope which is red-shifted, peaking at
36 \kms\ with respect to the LSR.

A 5$\times$3 grid with 20-arcsec spacing of 5\too 4 spectra orientated
with the long axis along the outflow was observed with the JCMT. None
of the spectra show significant emission (to a noise level of 85 mK in
a 1.8 \kms -wide channel), although a map of the integrated intensity
hints at a peak at the central position. Deeper integrations (to a
noise level of 24 mK in a 1.8 \kms -wide channel) were made at the map
centre and at the two CO peaks in the north-east lobe.  The 5\too 4
line was detected at the centre and the far CO peak at (45,15) arcsec
offset (with an antenna temperature of $\sim$0.2 K). The 2\too 1 SiO
data also show a peak at (45,15) arcsec offset from G35.2N, although
it is only at the 3-$\sigma$ level in Fig.~\ref{fig:sio21}. The 5\too
4 line was barely detected at the strongest blueshifted CO peak; thus
it would seem that CO intensity is not necessarily a good indicator of
where to search for SiO emission.

The 5\too 4 detections peak at different velocities: --0.5 \kms\ 
relative to the rest velocity at the centre and --7.5 \kms\ at (45,15)
arcsec offset. The width of each line is $\sim$8 \kms\ (0,0) and
$\sim$12 \kms\ (45,15). The 2\too 1 SiO emission does not appear to be
as broad as that from the 5\too 4 line.

\begin{figure*}
\centering 
\includegraphics[width=15cm]{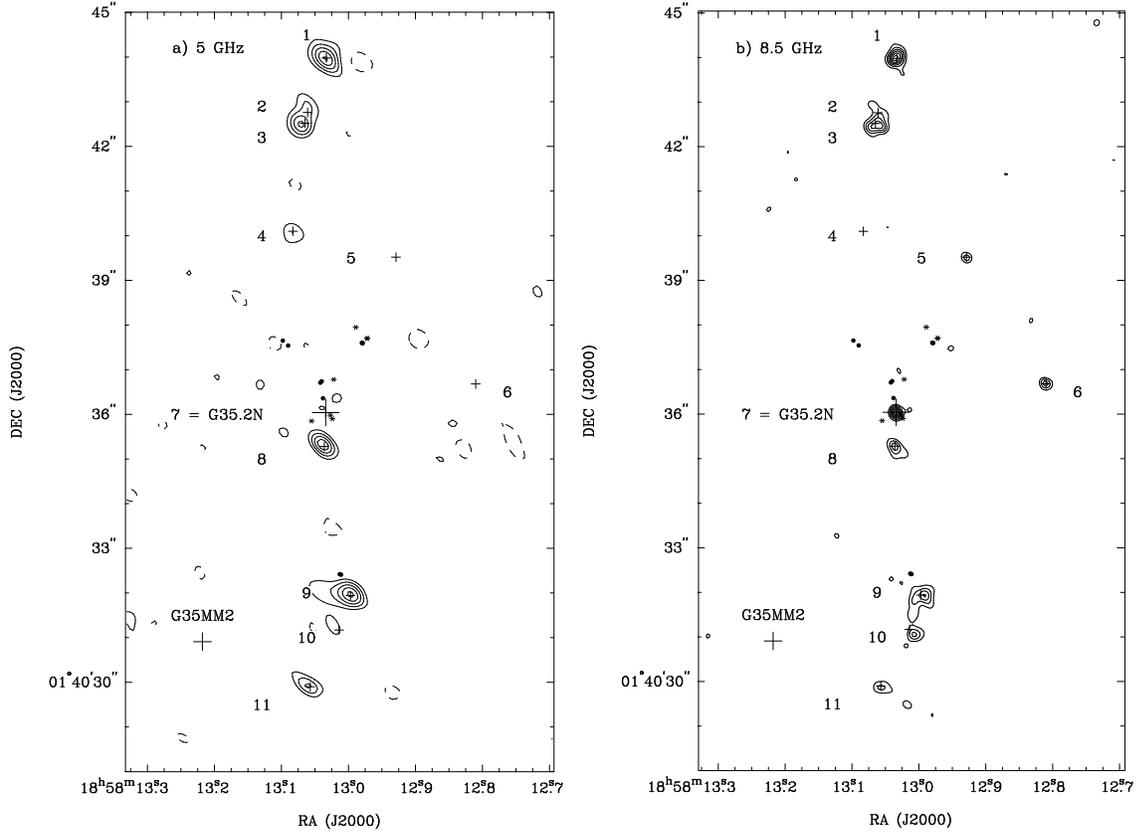}
\caption{a) 5-GHz radio continuum emission from G35.2N. The contours
  are at --5,3,5,7,9,11,13,15$\times$60$\mu$Jy per beam. The beam is
  0.57$\times$0.37 arcsec$^2$ at a position angle of 49$^\circ$. b)
  8.5-GHz radio continuum emission from G35.2N. The contours are at
  --5,3,5,7,9,11,13,15$\times$33$\mu$Jy per beam. The beam is
  0.25$\times$0.22 arcsec$^2$ at a position angle of 33$^\circ$.  In
  both images the positions of OH and H$_2$O masers from Hutawarakorn
  \& Cohen (1999) and Forster \& Caswell (1999) are marked by
  asterisks and filled circles respectively.  The radio sources listed
  in Table~\ref{tab:radio1} are marked by small crosses. The
  millimetre source, G35MM2, is also marked with a cross.
\label{fig:radio}}
\end{figure*}

\begin{table*}
\centering
\caption{Radio source parameters. The position offsets are in arcsec from
G35.2N, located at 
RA(J2000)=18$^{\rm h}$58$^{\rm m}$13.02$^{\rm s}$,
Dec(J2000)=01$^\circ$40$'$36.20$''$
Spectral indices are derived from
the fluxes given in columns 7 and 8, measured from the tapered images
(see text). The uncertainty in the peak/total fluxes is 0.06 mJy per
beam/0.15 mJy at 5 GHz and 25 $\mu$Jy per beam/50 $\mu$Jy at 8.5 GHz.
In the tapered maps the uncertainties in the total fluxes listed are
typically 0.1--0.2 mJy.
\label{tab:radio1}}
\begin{tabular}{ccccccccc}
\hline
Source & Offset   & \multicolumn{2}{c}{5-GHz flux} &
\multicolumn{2}{c}{8.5-GHz flux} & \multicolumn{2}{c}{Tapered flux} & Spectral \\
       & (arcsec)  & Peak & Total & Peak & Total & 5-GHz & 8.5-GHz & index \\
\hline
 1 & (0.0,7.8)     & 0.66    & 1.0    & 0.45    & 1.0    & 
1.0 & 0.9 & --0.2$\pm$0.4   \\
 2 & (0.4,6.5)     & 0.40    & 0.7    & 0.22    & 0.5    & 
 -- &  -- & -- \\
 3 & (0.5,6.4)     & 0.56    & 0.8    & 0.34    & 1.1    & 
0.9 & 1.4 & 0.8$\pm$0.4 \\
 4 & (0.7,3.9)     & 0.30    & 0.4    & $<$0.05 & $<$0.1 & 
0.2 & $<$0.1 & --1.3$\pm$0.4 \\
 5 & (--1.5,3.4)   & $<$0.13 & $<$0.2 & 0.20    & 0.2    & 
$<$0.1 & 0.2 & $>$1.3 \\
 6 & (--3.3,0.5)   & $<$0.13 & $<$0.1 & 0.28    & 0.3    & 
 -- &  -- & -- \\
 7 & (0,0)         & $<$0.20 & $<$0.2 & 0.48    & 0.5    & 
$<$0.2 & 0.4 & $>$1.3$\pm$1.4 \\
 8 & (0.1,--0.9)   & 0.66    & 0.6    & 0.25    & 0.6    & 
0.3 & 0.4 & 0.5$\pm$1.1 \\
 9 & (--0.4,--4.2) & 0.70    & 1.0    & 0.25    & 1.2    & 
1.3 & 1.0 & --0.5$\pm$0.3 \\ 
10 & (--0.3,--5.0) & 0.26    & 0.3    & 0.23    & 0.5    & 
0.3 & 0.4 & 0.5$\pm$1.1 \\ 
11 & (0.4,--6.2)   & 0.44    & 0.5    & 0.18    & 0.4    & 
0.6 & 0.5 & --0.3$\pm$0.7 \\ 
\hline
\end{tabular}
\end{table*}

\begin{figure*}
\centering 
\includegraphics[width=15cm]{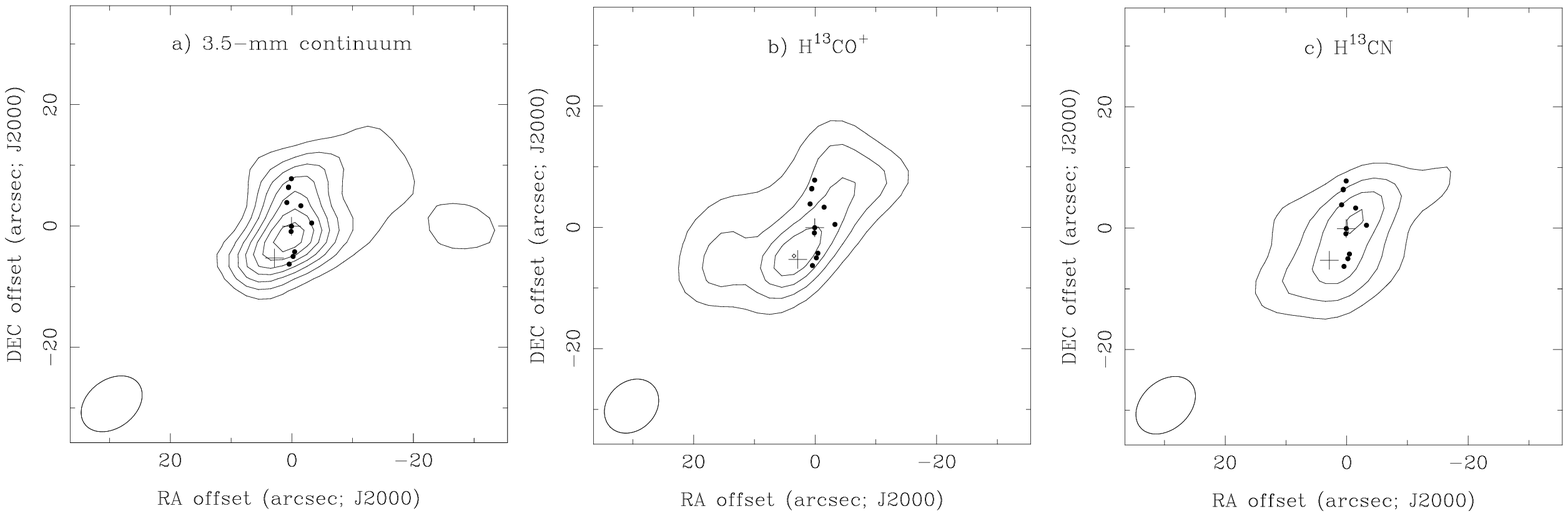}
\caption{a) BIMA 3.5-mm continuum emission (made with
  natural weighting). The peak flux is 82 mJy\,beam$^{-1}$ and the
  beam is 11.1$\times$7.9 arcsec$^2$ at --53$^\circ$. b)
  H$^{13}$CO$^+$ 1\too 0 and c) H$^{13}$CN 1\too 0 average intensity
  over the velocity range 32 to 36 \kms\ (covering the main $F$=2\too
  1 hyperfine of H$^{13}$CN).  In each panel the contours are spaced
  by -3, 3, 5, 7, 9, 11, 13, 15 times the noise level (8 mJy per beam,
  0.3 K and 0.3 K respectively). The radio sources are marked by
  filled circles with G35.2N and G35MM2 marked by crosses. Each
  image is centred on G35.2N.
\label{fig:cont1}}
\end{figure*}

\subsection{The radio jet}

Fig.~\ref{fig:radio} shows the radio continuum images of G35.2N at 5
and 8.5 GHz. The data are broadly consistent with HL88 in that they
show three main groups of components, one each to the north and south
and one corresponding to the source itself. However, the higher
resolution reveals structures not evident in the work of HL88. A total
of 11 radio sources can be identified in the two images (see
Table~\ref{tab:radio1}), although not all appear in both. The OH
masers of Hutawarakorn \& Cohen (1999) are distributed in a mostly
linear geometry parallel to the large-scale envelope, and are shown by
asterisks in Fig.~\ref{fig:radio}. Source \#7 in Fig.~\ref{fig:radio}
lies close to several OH and H$_2$O masers and is most likely to be
the driving source of the jet, which we refer to as G35.2N. The most
striking result is that G35.2N itself is not detected at 5 GHz.

The 8.5-GHz image reveals that the weak component $\sim$3.5 arcsec to
the west of the central source seen in the 15-GHz image of HL88 is
genuine (source \#6). A new point source (source \#5) is detected 3.7
arcsec north-west of G35.2N, also not seen in the 5-GHz map.
Fig.~\ref{fig:radio} also has the positions of H$_2$O masers taken
from Forster \& Caswell (1999). The majority of the OH masers are
associated with the central source and are arranged along a line
parallel with the major axis of the envelope (see below). Two of the
water masers are associated with the southern group of radio
components. They appear to lie upstream of the radio emission,
suggesting perhaps that they arise in the dense molecular gas behind
the shock. A study of their proper motions would confirm their
association with the radio components. The clear association of water
masers with shocks in outflows has only been seen before in two
sources (Claussen et al. 1998; Wilner et al. 1999).

The size of the field in Fig.~\ref{fig:radio} is comparable with the
15-arcsec JCMT beam, but contains all of the radio sources. No radio
emission is seen further away. The distribution of the radio
components on the sky strongly suggests that they represent knots in a
jet and that the jet is precessing. However, the implied half-angle of
the cone defined by the precession is less than 10 degrees, which
reinforces our earlier conclusion that the radio jet is probably not
responsible for driving the large-scale CO outflow seen in
Fig.~\ref{fig:co32int}.

\subsection{The G35.2N envelope}

The rotating clumpy envelope in which G35.2N is embedded was first
delineated in ammonia by Little et al. (1985), and later by Brebner et
al. (1987). Our 3.5 mm naturally-weighted continuum image
(Fig.~\ref{fig:cont1}a) has the same centrally-peaked structure seen
in the submillimetre dust continuum images of Dent et al. (1989) and
Vall\'ee \& Bastien (2000).  The deconvolved dimensions from a
two-dimensional gaussian fit are 22$\times$13 arcsec$^2$
(0.21$\times$0.13 pc$^2$ for a source at 2 kpc) at a position angle of
42 degrees west of north, consistent with the above dust observations.
The similar appearance of the 3.5-mm continuum and the submillimetre
dust emission suggests that our BIMA observations are tracing cool,
extended dust emission. The 3.5-mm continuum image extends along the
direction of the northern jet components, and even further north-west
out to encompass the north-west ammonia peak in the maps of Brebner et
al. (1987) and Little et al. (1985).

Observations of G35.2N made with OVRO (J.M. Carpenter, private
communication) reveal the presence of a second dust emission peak,
coincident with the main NH$_3$ (3,3) peak of Brebner et al. (1987)
and close to the SiO 2\too 1 peak in Fig.~\ref{fig:sio21}. Carpenter
has labelled this source G35MM2. Unfortunately our data do not have
the spatial resolution to confirm this but we do find that our \htcn\ 
and \htcop\ line emission does peak at the position of G35MM2 (see
below). However, G35MM2 has no radio counterpart in
Fig.~\ref{fig:radio} to a 4$\sigma$ level of 0.1 mJy per beam at 8.5
GHz.

The H$^{13}$CO$^+$ and H$^{13}$CN 1\too 0 integrated intensity maps
are shown in Fig.~\ref{fig:cont1}. The \htcop\ integrated emission
peaks on the (3,3) ammonia peak of Brebner et al.  (1987), coincident
with G35MM2 while the \htcn\ peaks on G35.2N. The emission from both
molecules exhibits the same north-west to south-east velocity gradient
observed in ammonia of Little et al. (1985). (It should be noted the
labelling of the channel maps of Little et al. are reversed.) The
velocity shift from north-west to south-east is $\sim$4 \kms.

\begin{figure*}
\centering 
\includegraphics[width=15cm]{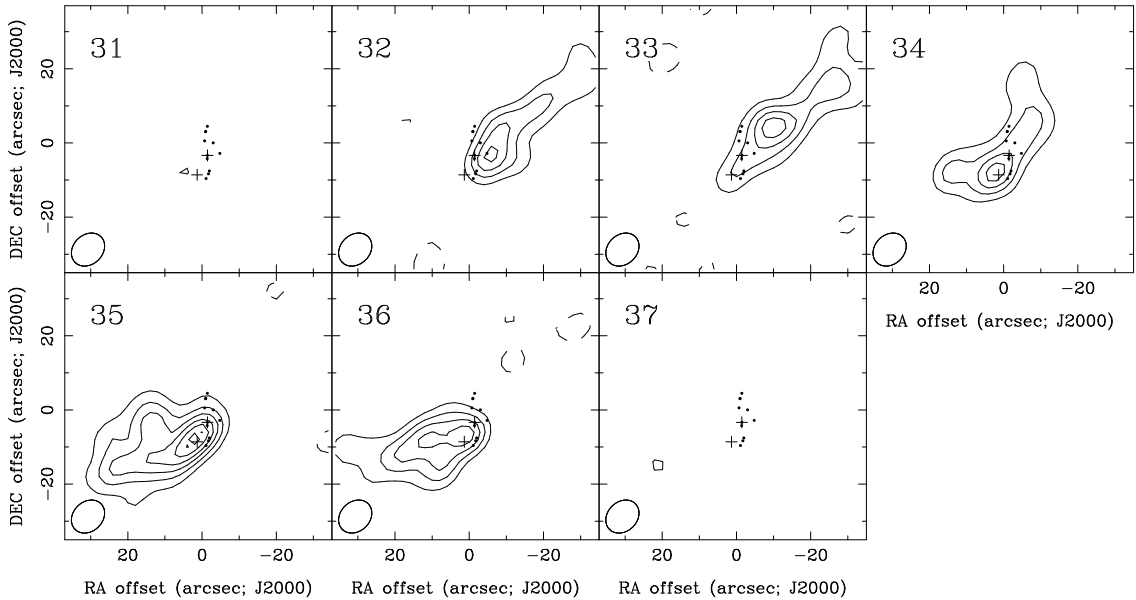}
\caption{ H$^{13}$CO$^+$ 1\too 0 emission integrated over five successive 
  1-\kms-wide velocity intervals from 31 to 37 \kms.  In each panel
  the contours are spaced by -3, 3, 5, 7, 9, 11, 13, 15 times the
  noise level ($\sim$0.3 K). The radio sources are marked by filled
  circles with G35.2N and G35MM2 marked by crosses. The image is
  centred on G35.2N. The {\sc clean} beam is 9.8$\times$7.9
  arcsec$^2$ at --46$^\circ$. The velocity of each panel is printed in
  the top left.
\label{fig:h13cop}}
\end{figure*}
\begin{figure*}
\centering 
\includegraphics[width=15cm]{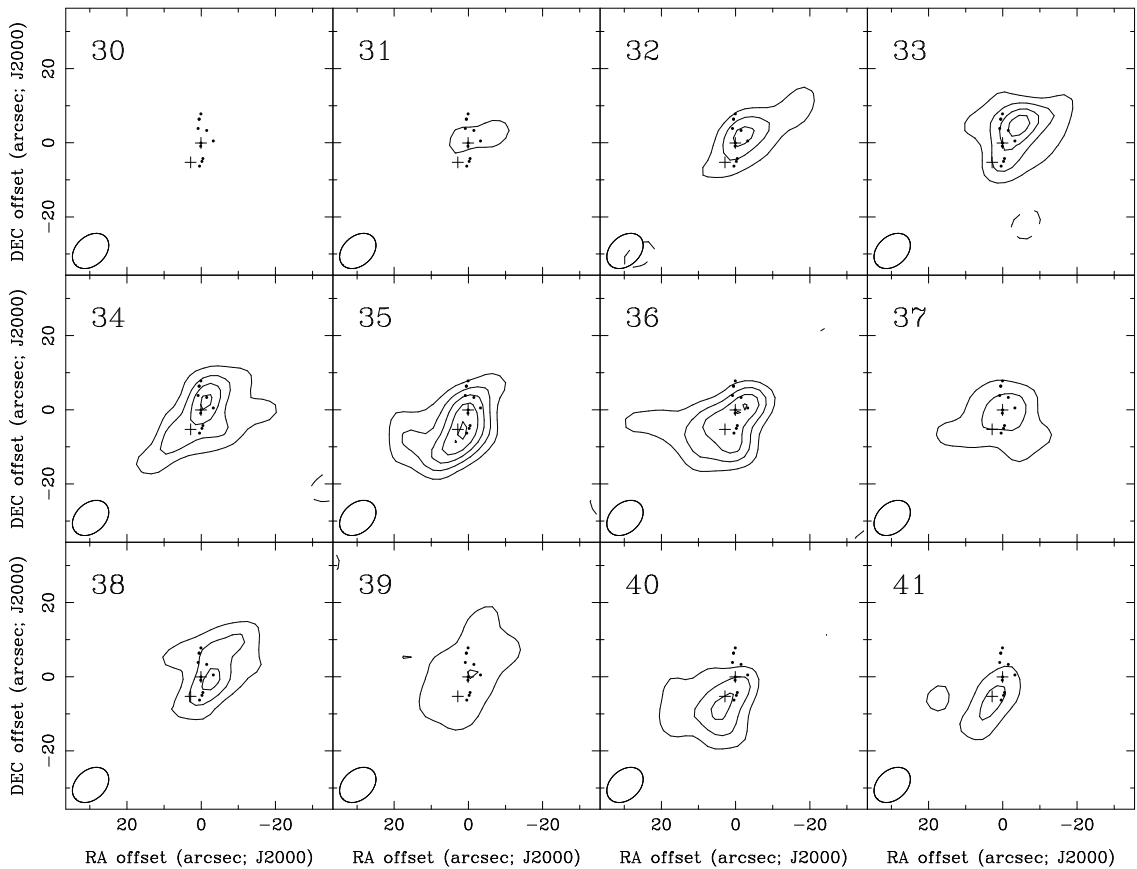}
\caption{H$^{13}$CN 1\too 0 emission integrated over successive 1-\kms-wide
  velocity intervals from 30 to 41 \kms. In each panel the contours
  are spaced by -3, 3, 5, 7, 9, 11, 13, 15 times the noise level
  ($\sim$0.3 K).The radio sources are marked by filled
  circles with G35.2N and G35MM2 marked by crosses. The image is
  centred on G35.2N.  The {\sc clean}
  beam is 11.1$\times$7.9 arcsec$^2$ at --48$^\circ$. The
  velocity of each panel is printed in the top left.
\label{fig:h13cn}}
\end{figure*}

Fig.~\ref{fig:h13cop} shows channel maps demonstrating the velocity
gradient across the envelope.  The envelope appears reasonably
continuous, although there is evidence of structure on scales smaller
than the $\sim$10-arcsec beam. The envelope is not resolved along the
minor axis. At the ambient velocity, the H$^{13}$CO$^+$ peaks on
G35MM2, although it extends north-west to include the location of
G35.2N itself. At 32 \kms\ the envelope extends 30 arcsec to the
north-west and shows a change in the position angle. This emission is
elongated and coincides with the weaker (3,3) ammonia peak of Brebner
et al. (1987). This peak is close to the SiO `jet' described above and
may represent the location of the driving source.

The brightness temperature of the \htcop\ emission is remarkably
uniform across the envelope at 2.5 to 3 K away from the central
position. The peak line brightness is 3.9 K towards G35MM2 where the
linewidth is 4.0 \kms; at the position of G35.2N the line is slightly
weaker at 3.5 K but broader at 5.2 \kms. A weak blue wing is seen in
both of these spectra which is absent further from the centre of the
core.

The H$^{13}$CN emission has qualitatively the same appearance as the
H$^{13}$CO$^+$ (Fig.~\ref{fig:h13cn}), although interpretation of the
velocity structure is complicated by the blending of the $F$=2\too 1
and $F$=1\too 1 hyperfine components which are separated by 4.86 \kms.
The H$^{13}$CN peaks more towards G35.2N, and also has a peak
brightness temperature of $\sim$3 K. Linewidths derived from Gaussian
fits to the H$^{13}$CN hyperfines are poorly constrained due to
blending of the hyperfine components but the linewidth peaks at the
position of G35.2N at $\sim$5.5--6.0 \kms, decreasing to 4.5--5.0 \kms
at the position of G35MM2. Further away from the centre, only the
stronger two hyperfines are detected with even lower linewidths
(around 2--2.5 \kms). 

Since the excitation of the 1\too 0 lines of \htcn\ and \htcop\ is
similar, it seems surprising that one is broader than the other.
Perhaps the greater linewidth derived from the H$^{13}$CN data implies
that these lines may be tracing more turbulent gas closer to the
embedded sources, or a component which is swept up in the outflows
from each source. Alternatively, since the \htcn\ beam is slightly
larger than that for \htcop\ it could be tracing more of the
systematic velocity gradient along the core. This seems unlikely since
the extra contribution to the \htcn\ linewidth from the velocity
gradient (13\,km\,s$^{-1}$\,pc$^{-1}$: see below) in the additional 1.3
arcsec may be estimated to be only $\sim$0.16 km\,s$^{-1}$. Finally,
the blending of two of the hyperfine components may be biasing the
Gaussian fits towards larger values for the linewidth.

The observed ratio of the \htcn\ hyperfine components is consistent
with optically thin emission from molecules in local thermodynamic
equilibrium (LTE) at all positions. An upper limit to the total
optical depth is $\sim$0.1 (equal to the sum of the optical depths of
each hyperfine component).

\section{Analysis}

\subsection{Outflow parameters}

While the data presented in Section 3 demonstrate that there are
probably several outflows, it is not obvious how they should be
separated and so the following analysis examines the whole of the
high-velocity emission.

The outflow parameters have been calculated using standard analysis
techniques (e.g. Bally \& Lada 1983), initially assuming optically
thin emission and that the level populations are characterized by a
single excitation temperature at all velocities. Minimum values for
the mass, momentum and kinetic energy can be derived assuming an
excitation temperature equal to the energy of the upper level
(Macdonald et al. 1996), which is $\sim$33 K for the $J$=3 level of
CO. A CO abundance of 10$^{-4}$ relative to H$_2$ was assumed. The
choice of the velocity range to exclude from the calculation is
determined in part by the amount of the line affected by the severe
self-absorption. We have chosen to ignore emission less than 4 \kms\ 
away from the line centre (in accord with Little et al. 1985).

If the optical depth, $\tau$, is non-negligible, then the mass and
outflow energetics will be underestimated by a factor of
$\tau$/(1--\,e$^{-\tau}$). We have estimated the optical depth by
comparing \twco\ and \thco\ 3\too 2 spectra recorded at the same
positions across the outflow (see e.g.  Lada 1985). We have assumed
the [\twco]/[\thco] ratio is 75. This analysis shows that over the
velocity range considered, the beam-averaged optical depth generally
falls monotonically from a value of $\sim$3--5 (at velocities 4 to 6
\kms\ from the line centre) to less than $\sim$1 at velocities greater
than about 10 \kms\ from the line centre. The signal-to-noise ratio is
not high enough to extend beyond 10 \kms. The mean value between 4 and
10 \kms\ from the line centre is $\sim$2--3. The fraction of material
lying at the lower velocities corresponds to about half the calculated
mass, which means that by assuming low optical depth (and that the
emission uniformly fills the beam), we are underestimating the total
outflow mass by about a factor of two. The momentum and kinetic energy
of the flow are less susceptible to the effects of finite optical
depth since they are weighted by velocity, and the highest velocity
material has the lowest beam-averaged optical depth.

From the results above we can fit the optical depth with a function
that falls off with velocity as a power law of the form $\tau(v)
\simeq 57/v^{1.7}$. This reproduces the three values quoted above to
within 10 per cent. We can use this to make a correction for optical
depth with velocity for the outflow since. Assuming such a power law,
the optical depth does not get below 0.1 until a velocity 45 \kms\ 
from the line centre is reached, a velocity which is equal to or
higher than the maximum velocity in each lobe.

\begin{table}
\caption{Outflow parameters. No correction for optical depth or
inclination has been applied to these values. No attempt has been
made to distinguish individual flows.
\label{tab:coparms}}
\centering
\begin{tabular}{lcc}
\hline
Parameter & Red lobe & Blue lobe \\
\hline
Mass (M$_\odot$) & 3.7 & 7.7 \\
Momentum (M$_\odot$~km\,s$^{-1}$) & 31.3 & 81.7 \\
Energy (erg) & 3.2(45) & 12.4(45) \\
Size (pc) & 0.37 & 0.37 \\
Mean velocity (km\,s$^{-1}$) & 8.3 & 10.6 \\
Timescale (yr) & 4.3(4) & 3.4(4) \\
Luminosity (L$_\odot$) & 0.59 & 2.9 \\
Force (M$_\odot$~km\,s$^{-1}$~yr$^{-1}$) & 7.3(--4) & 2.4(--3) \\
\hline
\end{tabular}
\end{table}

Table~\ref{tab:coparms} lists the relevant outflow parameters with the
above optical depth correction applied. The size is the
intensity-weighted mean distance from the central source. The mean
velocity is calculated from the ratio of the momentum to the mass. The
luminosity is the product of the energy and the dynamical timescale,
while the force is the momentum divided by the dynamical timescale. In
addition, no attempt has been made to separate the two outflows,
although the jet-flow is likely to contribute only a small amount to
the total.

Comparison of our results with the 1\too 0 results of Little et al.
(1985) reveals a large discrepancy. For example, we calculate the
outflow mass as $\sim$11 M$_\odot$ here compared with 30 M$_\odot$
(correcting the value of Little et al. for the different CO
abundance). Excitation effects probably account for the large
discrepancy between the two values. The 3\too 2 line is excited in
denser gas, and thus its emission will arise from a smaller area. It
is unlikely that assuming 30 K as the excitation temperature
contributes significantly to the difference in 1\too 0 and 3\too 2 CO
mass estimates as the fraction of molecules in the $J$=3 level
decreases by only 50 per cent if we choose a temperature as low as 15
K, or as high as 100 K. The observed ratio of the 3\too 2 and 4\too 3
lines unfortunately places little constraint on the excitation
temperature. The ratio of the mean spectra has a roughly constant
value of 1.1$\pm$0.2 at all velocities where the sensitivity is high
enough, consistent with our assumption of 33 K (assuming that the gas
uniformly fills both beams). There is also a weak trend for the
higher-velocity material to have a higher excitation temperature.

\begin{figure*}
\centering 
\includegraphics[width=17cm]{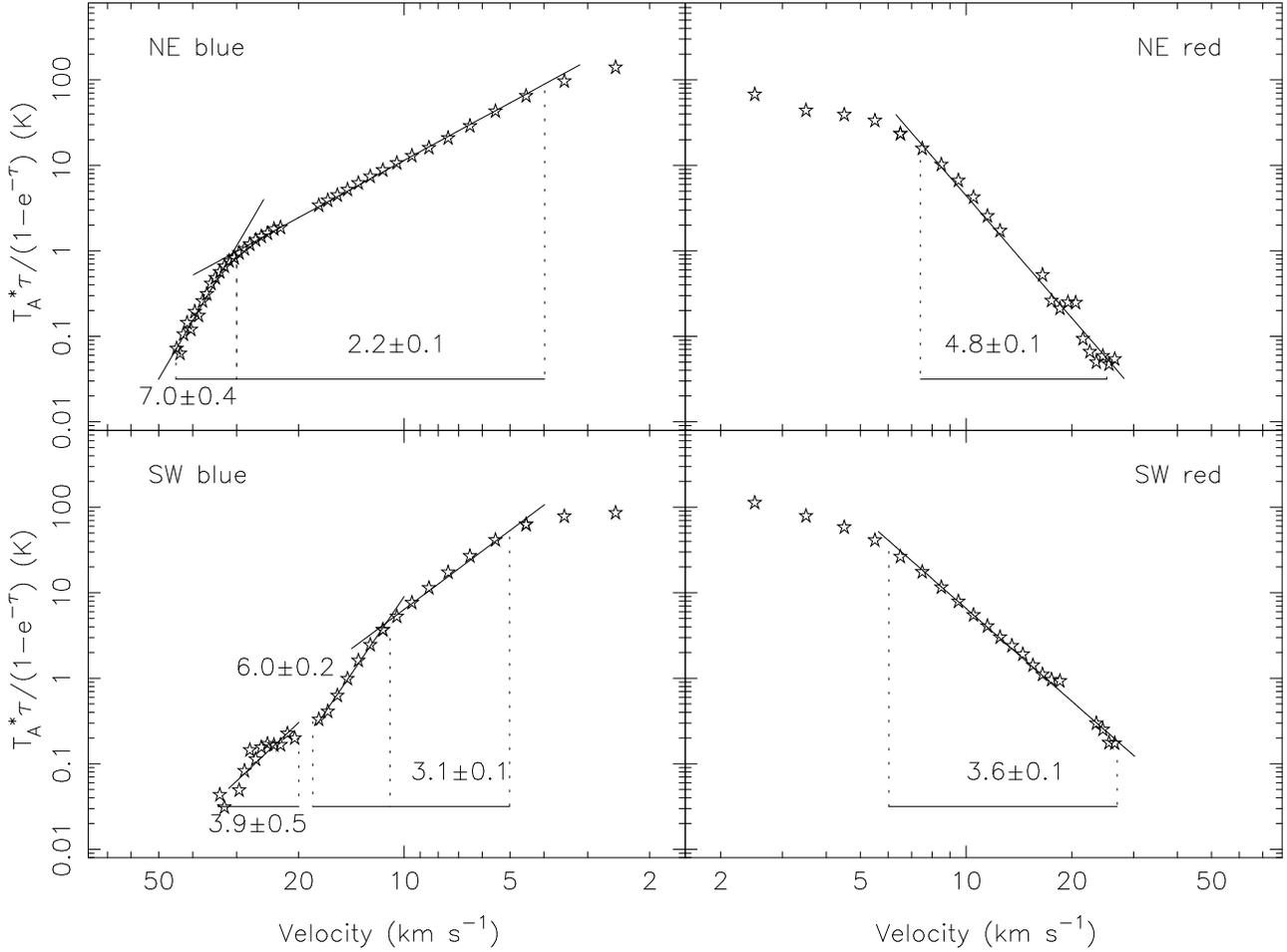}
\caption{Mass-velocity plots for the red- and blueshifted gas in both
  the north-east and south-west lobes. One or more power-laws have
  been fitted to the data, and the index is shown in each panel. a)
  Velocity ranging from 5--45 km\,s$^{-1}$ from the rest velocity (34
  \kms) b) 6--25 \kms, c) 6--34 \kms, and d) 7--27 \kms. The gaps at
  $\sim$15--20 \kms\ where data have been removed from the fit
  probably represent an unrelated molecular cloud 
  along the line of sight.
\label{fig:mvplot}}
\end{figure*}

\begin{table}
\centering
\caption{Listing of $\gamma$ as derived from fits to each lobe, both
  for uncorrected (`raw') data and optical depth-corrected data
  (`corr'). The 
  fits are plotted on Fig.~\ref{fig:mvplot} over the ranges given in
  this table. \label{tab:gamma}}
\begin{tabular}{lcccc}
\hline
Lobe & Velocity Range & $\gamma_{\rm raw}$ & Velocity Range & $\gamma_{\rm corr}$ \\
     & (\kms) & & (\kms) & \\
\hline
NE blue & ~5--10 & 1.0$\pm$0.1 & ~5--30 & 2.2$\pm$0.1 \\
  ~     & 10--30 & 2.0$\pm$0.1 &  & \\
  ~     & 30--44 & 6.6$\pm$0.4 & 30--44 & 7.0$\pm$0.4 \\
NE red  & ~9--25 & 4.9$\pm$0.4 & ~6--25 & 4.8$\pm$0.4 \\
SW blue & ~6--11 & 2.4$\pm$0.1 & ~6--11 & 3.1$\pm$0.1 \\
  ~     & 11--18 & 5.5$\pm$0.2 & 11--18 & 6.0$\pm$0.2 \\
  ~     & 20--34 & 3.2$\pm$0.5 & 20--34 & 3.9$\pm$0.5 \\
SW red  & ~7--27 & 2.8$\pm$0.1 & ~7--27 & 3.6$\pm$0.1 \\
\hline
\end{tabular}
\end{table}

We have also calculated (and plotted in Fig.~\ref{fig:mvplot}) the
mass as a function of velocity for each lobe to investigate its
usefulness in constraining models of outflows. We have again assumed
that the emission is characterized by a constant excitation
temperature and uniformly fills the beam at all velocities, so that
the outflow mass as a function of velocity is directly proportional to
the observed brightness temperature (summed over all spectra in the
lobe) at that velocity corrected for optical depth.  We have fitted
power laws of the form $v^{-\gamma}$ to the data. We find that not
only are these power laws different from the $v^{-1.8}$ found by
Masson \& Chernin (1992), but that a single power law does not always
reproduce the variation of mass over the whole velocity range. In both
the red- and blue-shifted gas there are velocity ranges which have
indices much greater than 2, most notably for the highest-velocity
blue-shifted gas which has a value for $\gamma$ as high as 7.0 in the
north-east lobe (Fig.~\ref{fig:mvplot}). However, it is remarkable
that with the optical depth correction applied, the north-east blue
lobe does have a value of $\gamma$ close to 2 which fits the data out
to 30 \kms\ from the line centre. The north-east blue lobe also shows
a turnover in $\gamma$ at the highest velocities, a feature seen in a
number of other outflows (Fich \& Lada 1997; others?). The other lobes
do not show such a turnover, but they also do not extend as far in
velocity.

\subsection{SiO excitation}

With the two transitions of SiO we have been able to apply a radiative
transfer code (which employs the large velocity gradient or LVG
approximation: Goldreich \& Kwan 1974) to try and constrain some of
the parameters of the shocked gas in the outflow. We used it in two
ways: in the first we made no correction for the difference in
beamsize, and in the second we constructed a 2\too 1 image with a
22-arcsec restoring beam from which we then obtained spectra.

We attempted to model the lines from the position of MM2, but we were
only able to place some limits on the gas parameters. The modelling
confirmed some general intuitive conclusions regarding the
excitation of SiO in the shocked gas; either a low density/high
abundance or high density/low abundance solution, with the temperature
remaining relatively low (30--50 K). The high-density
($\sim$few$\times$10$^6$ cm$^{-3}$) solutions were found at the lower
temperatures.

The main difficulty with deriving the gas parameters from the SiO data
is that the emitting area is very much smaller than the 22-arcsec beam
(and even the 10-arcsec BIMA beam). The models of Schilke et
al. (1997) predict an SiO-emitting region of order 10$^{16}$ cm in
length, which corresponds to 0.33 arcsec at the distance to G35.2N (2
kpc). This means that the beam-averaged parameters generally are not
good estimates of the true parameters, and probably underestimate the
true density in the SiO emitting regions by an order of magnitude or
more. 

Are we missing any extended SiO emission? Codella, Bachiller \&
Reipurth (1999) show that towards some outflow sources the lower $J$
SiO emission (2\too 1 and 3\too 2) has a diffuse extended component
compared with the 5\too 4 emission. In addition, interferometric
observations of HH7--11 show that only 10 per cent of the SiO flux is
recovered (Bachiller et al. 1998). If the SiO 2\too 1 emission
associated with the G35.2N region is extended then our BIMA
observations may be missing a significant fraction, leading to
erroneous conclusions regarding its distribution and excitation. It is
difficult to know whether this is the case without single-dish
observations of the 2\too 1 SiO line. The effect of missing flux on
our LVG modelling will be to lower the 5\too 4/2\too 1 ratio, most
likely leading to solutions which favour lower density and/or
temperature.

\subsection{The properties of the envelope}

A two-dimensional gaussian fit to the 3.5-mm continuum emission yields
source dimensions of 22$\times$13 arcsec$^2$ at a position angle
of --42 degrees. The peak flux in the model is 71 mJy\,beam$^{-1}$ and
the total flux is 241 mJy. If the radio sources have the spectral
indices listed in Table~\ref{tab:radio1} then they could contribute up
to 20 mJy at 86 GHz. This is roughly 10 per cent of the total, but
since most of the radio emission is confined to a single BIMA beam, it
is possible that the emission from the wind contributes a significant
fraction (30 per cent) to the peak flux at the position of
G35.2N. Observations at intermediate frequencies and higher resolution
millimetre observations are necessary to determine the exact
contributions of the free-free emission.

Assuming a dust temperature of 30 K (Dent et al. 1989), we derive
masses for the whole envelope of between 600 and 1400 M$_\odot$ from
the 3.5-mm continuum data, depending on the value and
frequency-dependence of the absorption coefficient (Hildebrand 1983;
Mezger et al. 1987).

\begin{figure}
\centering
\includegraphics[width=8cm]{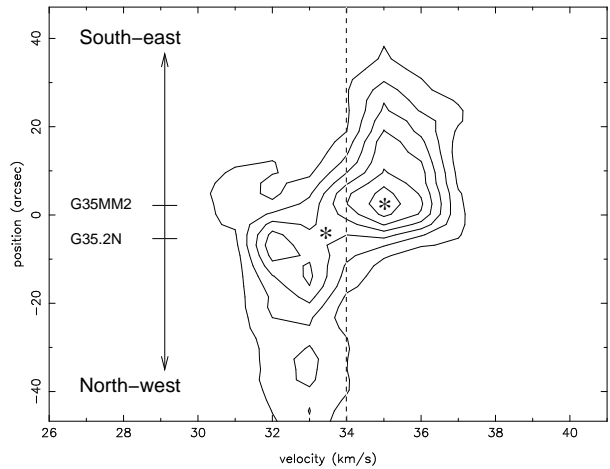}
\caption{Position-velocity diagram for the \htcop\ emission. The
  $y$-axis represents offsets in arcsec along the cut which has a
  position angle of 55 degrees west of north. The positions along the
  cut of G35.2N and G35MM2 are labelled and marked by asterisks.
\label{fig:pvdiag}}
\end{figure}

The \htcop\ emission shows a velocity gradient along the long axis of
the envelope of 13 \kms \,pc$^{-1}$, which translates to a rotation
period of 75000 years. A position-velocity cut along the long axis
(shown in Fig.~\ref{fig:pvdiag}) shows that the velocity gradient is
not uniformly smooth, but appears to show evidence for three velocity
components within the envelope, one at 32.5 \kms\ (to the north-west),
one at the systemic velocty of 34 \kms\ and one at 35 \kms\ (to the
south-east). Thus the envelope is not continuous but fragmented into a
number of cores each harbouring one or more YSOs.  Alternatively the
\htcop\ may be tracing components of the outflow, blue-shifted in the
north-west and red-shifted in the south-east.  However, this may be
true only for the extreme red- and blue-shifted \htcop\ emission since
the same velocity gradient is observed in the ammonia data of Little
et al. (1985).

Virial masses may be derived from the linewidths given above. For a
uniform sphere, the virial mass is given by $M_{\rm vir} = 209 \Delta
v^2 r$\,M$_\odot$, where $\Delta v$ is the linewidth in \kms\ and $r$
is the radius in pc (taken to be the appropriate beam radius). At the
position of G35.2N, the virial mass is estimated to be
$\sim$350\,M$_\odot$, decreasing to $\sim$210\,M$_\odot$ at the
position of G35MM2. These masses are larger than those derived from
the ammonia observations of Little et al. (1985) which suggests that
the \htcn\ may be tracing regions with enhanced velocity dispersion,
probably the outflow.

From the brightness of the \htcn\ lines and assuming a total optical
depth of 0.1 (see above) we derive excitation temperatures of 50 and
45 K at the positions of G35.2N and G35MM2 respectively, assuming the
emission uniformly fills the beam. These are higher than previous
estimates (typically 30 K; Little et al.  1987), consistent with
tracing warm material close to the embedded sources. Assuming
optically thin LTE emission, a column density of
1.6$\times$10$^{14}$\,cm$^{-2}$ is derived for \htcn\ towards G35.2N.
The virial mass above translates to a molecular hydrogen column
density of 3.6$\times$10$^{24}$\,cm$^{-2}$, which implies an abundance
of 4.4$\times$10$^{-11}$ for \htcn. The \htcn\ abundance is not well
determined but using an abundance for the main isotopomer of
2$\times$10$^{-8}$ (Irvine, Goldsmith \& Hjalmarson 1987) and assuming
a $^{12}$C/$^{13}$C ratio of 75 a value of 2.7$\times$10$^{-10}$ may
be derived.  An alternative estimate for the \htcn\ abundance may be
derived by comparing the dust mass above with the estimate for the
envelope from the \htcn\ lines. This method give a value of
$\sim$1.1$\times$10$^{-10}$ with a 50 per cent spread in values. These
calculated values are factors of 2 to 5 smaller than that derived from
the value given by Irvine et al. (1987). This may indicate that HCN is
freezing out onto dust grains near the centre of the core, but given
the large uncertainties in both the H$_2$ column density and the
`standard' value for the \htcn\ abundance we regard this conclusion as
tentative at best.

Since the \htcop\ is not tracing gas as close to the embedded sources
as the \htcn\ (see section 3.5), the excitation temperature for the
\htcop\ will be lower. We will assume a value of 30 K as appears to
be appropriate for the less dense gas. A \htcop\ column density of
4.0$\times$10$^{13}$\,cm$^{-2}$ is derived at the centre of the
envelope.

\subsection{The nature of the radio components}

The majority of the radio components detected (sources \#1--4 and
8--11) lie along a line which suggest they are part of a collimated
jet driven by G35.2N. There are two which probably are not and which
may be individual YSOs or extragalactic in nature. Since these are
seen at 8.5 GHz but not at 5 GHz they have a positive spectral index
and therefore are most likely to be YSOs. Source \#6 lies between two
of the mid-infrared sources detected by Fuller et al. (2001), and was
suggested by these authors to be responsible for a separate outflow.
Within the small field imaged ($<20$ arcsec diameter) we may expect to
detect $<$1 extragalactic source at 8.5 GHz (Anglada et al. 1999).

We have calculated spectral indices (assuming total flux density
$\propto \nu^\alpha$) for all the components which show up at both 5
and 8.5 GHz and placed limits on those which are visible at only a
single frequency. To avoid problems with differing beamsizes, both
maps were tapered to give beams with similar half-power dimensions of
0.65$\times$0.56 arcsec. The required taper was 300 k$\lambda$ at 5
GHz, and 220 k$\lambda$ at 8.5 GHz. The noise in each image was
similar at 75--80 $\mu$Jy per beam. The spectral indices are listed in
Table~\ref{tab:radio1}. Those sources without spectral indices were
only present in the full-resolution images.

The uncertainties in the calculated spectral indices are significant
and in most cases it is difficult to unambiguously determine the
emission mechanism responsible for each component.  Sources \#5, \#6
and \#7 have clearly positive spectral indices, and therefore probably
mark the location of embedded luminous YSOs. In particular \#7 is the
source of the jet due to its location with respect to the OH and water
masers in the region (Fig~\ref{fig:radio}). For the remaining
components it seems likely that the spectral indices are consistent
with optically thin thermal free-free or weak gyrosynchrotron
emission.

Models of radio emission from shocks can account for both negative and
positive spectral indices, but not with a single model (Henriksen et
al. 1991; Ghavamian \& Hartigan 1998). In G35.2N, therefore, the
mechanism must be different in each case and it suggests that the
shocks in outflows from massive YSOs contain weakly relativistic
electrons. In this respect, G35.2N differs from the other well-known
radio jet sources such as HH80--81 (Mart\'{\i} et al. 1993), W3-H$_2$O
(Wilner et al. 1999) and Serpens (Rodr\'{\i}guez et al. 1989). All
these sources appear to have symmetric jets with negative (and equal)
spectral indices for components in either direction from the central
source.

\subsubsection{Extended radio emission: an ionized cavity}

It is instructive to compare the 5-GHz flux in our A-configuration map
with the fluxes derived from the lower-resolution observations of
Dent, Little \& White (1984) and HL88. From such a comparison we
conclude that there is evidence for an evacuated cavity which contains
ionized gas. This cavity is associated with the jet source, G35.2N.
Furthermore it may be that weak emission has been detected along the
path of the beam of the jet itself.

Dent et al. (1984) derive a total of $\sim$14 mJy for their
double-peaked structure. An extended (`diffuse') component is observed
in their data, contributing about 40 per cent of the total flux
density. HL88 measure a total flux density of $\sim$8 mJy, about 60
per cent of that of Dent et al. (1984). These authors deduce that a
gaussian source with FWHM dimensions 10$\times$2 arcsec$^2$ can
account for the difference between the two measurements. Such a source
may represent extended emission from ionized gas within a cavity swept
out by the jet and/or a stellar wind.

The total 5-GHz flux density in our data is $\sim$5.4 mJy, or 68 per
cent of the HL88 flux, suggesting that our A-configuration
observations are further resolving out the emission. The flux missing
from our 5-GHz image is probably from the extended feature seen in the
HL88 image which links the northern group of components with the
central region. This may be ionized material within the jet itself or
at the shocked boundaries of the jet and ambient cloud.

\section{Discussion}

\subsection{Precessing outflows in G35.2--0.7N?}

It has long been suspected that the driving source for the CO flow is
precessing (HL88, Little et al. 1998). The 3\too 2 CO data appear to
support this interpretation. The blue lobe in particular
(Fig.~\ref{fig:co32int}a) has an `S'-shape similar to that expected
for a precessing flow (e.g. Cliffe, Frank \& Jones 1996). The problem
with this interpretation has always been the large offset in position
angle between the radio jet (roughly north-south) and the CO flow (58
degrees east of north). 

The data presented in this paper help to resolve this issue, showing
that there are perhaps as many as four outflows, all of which
contribute to the overall CO appearance: 1) the radio jet and CO flow
from G35.2N; 2) the CO flow from G35MM2; 3) the SiO flow; and possibly
4) the flow driven by radio source \#6 (only observed indirectly
through the infrared observations of Fuller et al. 2001).

The CO flow associated with the radio jet is primarily identified from
the finger of redshifted CO emission seen in Fig.~\ref{fig:co32int},
although it is also evident at velocities close to the systemic
velocity (see the 32.5 \kms -panel in Fig.~\ref{fig:co32chan}), which
suggests that it lies close to the plane of the sky. However, the fact
that the infrared reflection nebula is brighter in the north suggests
that it is tilted towards us, so that CO emission from the jet outflow
should be blueshifted. In general the assumption that a bright
reflection nebula indicates an region inclined towards us is based on
the cloud in which the nebula is embedded being uniform. If, say, the
G35.2N cloud is denser to the south and west then the southern lobe of
a reflection nebula will be more heavily extinguished compared with
the northern lobe, especially if the outflow is close to the plane of
the sky. Such a geometry would yield a brighter northern reflection
nebula despite being tilted away from us provided the inclination to
the plane of the sky was not significant as appears to be the case
here.

The superposition of multiple flows with slightly different position
angles can account for the impression of a single precessing flow.
Therefore a precession interpretation seems unnecessary. However, it
is likely that the driving sources of at least two of the flows are
themselves precessing. The radio data {\em do} appear to show clear
evidence for a precessing jet, with the individual knots of radio
emission showing offsets of upto an arcsecond from a projected
straight line through the central source and curving away at each end
of the jet (Fig.~\ref{fig:radio}). The peaks in the CO flow
(Fig.~\ref{fig:co32int}) do not lie along a straight line which
suggests that the flow from G35MM2 is also precessing.

\subsection{The mass-velocity relation}

The question of whether the mass-velocity index ($\gamma$, where $m_v
\propto v^{-\gamma}$) places a useful constraint on outflow models
remains unanswered. Masson \& Chernin (1992) derived a value of 1.8
for several flows, while a number of recent observational studies have
shown that mass at high-velocity decreases more rapidly than as
$v^{-2}$ (Davis et al. 1998, Gibb \& Davis 1998; Shepherd et al. 1998;
Fich \& Lada 1997). The model of Matzner \& McKee (1999) predicts that
the index should indeed be close to 2, with little sensitivity to the
environment. Numerical studies of jets are less clear-cut and while it
is possible to obtain large indices, they suggest that the environment
of the protostar and outflow does play an important role (Downes \&
Ray 1999).

More recent work by Lee et al. (2001) confirms that jet-driven
outflows do tend to have steeper mass-velocity relations ($\gamma \sim
3.5$ can be achieved) than similar wind-driven outflow models ($\gamma
\simeq 1.8$). However, their wind-driven models also exhibit another
feature observed in many outflows (including G35.2N) and not observed
in the jet model: a break in the power law where $\gamma$ changes from
a relatively low value (say 2) increasing to a much steeper value (4
or greater). The presence of a high value for $\gamma$ as well as a
break in the power law may suggest that a wide-angle wind is necessary
in addition to a jet. In the models of Lee et al. (2001), the break
point occurs at lower velocities for outflows lying closer to the
plane of the sky, demonstrating the contraction of the velocity field
expected for a predominantly forward-directed flow.

Recently, Arce \& Goodman (2001a) have suggested that steep
mass-velocity plots can be caused by the superposition of several
mass-velocity curves for an episodic outflow. In their picture the
individual outbursts have $\gamma$=2, in line with the
momentum-conserving models of Matzner \& McKee (1999), but vary in
maximum velocity and amount of mass swept up in each outburst. A
similar effect could conceivably result from the superposition of a
number of flows with differing properties, such as is the case here in
G35.2N. Higher resolution CO observations would allow us to test this
as we would be able to resolve the individual outflows and derive
their individual mass-velocity indices.

Can the breakdown in our assumptions also explain the observed
behaviour?  All of our assumptions cause us to underestimate the mass
in the flow. The net effect of correcting for all of our main
assumptions is to steepen the mass-velocity variation close to the
line centre, and flatten it at high velocities, perhaps enough to
allow a single power law to fit the data. The effect of correcting for
finite optical depth is shown in Table~\ref{tab:gamma} where higher
values of $\gamma$ are derived, a consequence of the fact that the
mass will be corrected upwards at low velocities (e.g. Arce \& Goodman
2001b). If the excitation temperature increases with velocity then our
isothermal assumption will underestimate the mass at high velocities,
although this will only a small effect as the population of the $J$=3
level changes by less than 50 per cent from 33 to 100 K.

If the beam-filling factor of the high-velocity gas decreases with
velocity, then correcting for this will also decrease the value of
$\gamma$. This effect is potentially much larger than either of the
previous two (although optical depth effects are important at low
velocity). It has some observational support as well in that in a
number of outflows the highest-velocity emission does appear to be
unresolved (see Fig.~\ref{fig:co32chan}; also Richer et al. 1992; Gibb
\& Heaton 1993; Fich \& Lada 1997), even in interferometric studies
(e.g. Gueth \& Guilloteau 1999). Incomplete beam filling may give rise
to a break in the mass-velocity spectrum. If we define $v_{\rm fill}$
as the highest velocity at which the emission fills the telescope
beam, then for velocities less than $v_{\rm fill}$, we expect that the
mass-velocity plot will have a power-law equal to the true value. For
velocities greater than $v_{\rm fill}$, the antenna temperature will
be reduced, leading to an underestimate of the mass at these
velocities and thus causing the mass-velocity plot to fall off more
quickly. If the beam-filling factor is a power law of the form
$v^{-\alpha_{\rm fill}}$ then a power-law fit to the observed
mass-velocity spectrum will have an effective $\gamma$ given by
$\alpha_{\rm fill} + \gamma$. The observed values for $\gamma$ given
in Table~\ref{tab:gamma} suggest that $\alpha_{\rm fill}$ may be as
high as 5. It would be useful to know how the filling factor varies
for a given beam size in models of outflows. For a sample of similar
outflows observed with a beam with FWHM $\theta$, we would expect more
distant outflows to have show a break in the $m$--$v$ plot at smaller
velocities.  Equivalently we would expect that a sample of flows
observed at the same distance would show the same break point, all
else being equal (excitation, energetics, inclination). A test of this
hypothesis may be through the observations of Ridge \& Moore (2001),
who did not observe any such correlation.

\subsection{Is G35MM2 a high-mass protostar?}

From the luminosity and the presence of the radio jet it is clear that
the G35.2N core houses massive YSOs (Dent et al. 1985, HL88).  Since
the envelope appears to be centred on and rotating about G35.2N, it
seems likely that it is the most massive object in the group. But what
of the other sources, including G35MM2? G35MM2 has no radio source
associated with it and is a weaker dust continuum source, indicating a
less massive object. Yet its location relative to the CO appears to
suggest that it is responsible for a significant portion of the
observed CO outflow. Perhaps it represents a lower-mass protostellar
object, whose radio emission would be undetectable at 2 kpc. This
seems plausible given the weak emission from the G35.2N itself
(Table~\ref{tab:radio1}). However, G35MM2 is probably not
significantly less massive than G35.2N because the mass of the
outflowing gas (at least 7 M$_\odot$) seems to rule out a low-mass
driving source (e.g.  Richer et al. 2000), unless a significant
fraction of the outflowing gas arises from another outflow driven by
one of the other YSOs in this region.

What does seem likely is that it is more deeply embedded than G35.2N
because the ammonia maps of Little et al. (1985) and Brebner et al.
(1987) all have clear peaks towards G35MM2, particularly in the case
of the (3,3) line (Brebner et al. 1987). In addition our
H$^{13}$CO$^+$ map is skewed towards the position of G35MM2, although
the \htcn\ map is not. (This is most likely due to the \htcn\ tracing
warmer gas towards G35.2N.) The virial mass calculated towards G35MM2
is high at 210\,M$_\odot$, which implies a large mean density of
8$\times$10$^6$\,cm$^{-3}$. The \htcop\ and \htcn\ and therefore
probably tracing a component from the outflow.  It seems plausible
that G35MM2 may be a massive YSO still in its collapse phase where its
H\,{\sc ii} region is quenched by the action of infalling gas
(Walmsley 1995). This is consistent with the scenario where a bipolar
outflow is intrinsically linked to active accretion.

Further higher-resolution observations are clearly necessary to
determine the properties of G35MM2 and the other sources around
G35.2N.

\subsection{An organized cluster of massive outflow sources}

A simple theory of star formation leads to the conclusion that a
rotating envelope which collapses will form a pancake-like core in
which new stars emerge with parallel rotation axes perpendicular to
the long axis of the core. In practice, such a simple idea has rarely
been observed, and many cores where multiple stars are forming show
essentially random orientations of the respective outflow axes both
with respect to one another and with respect to the parent core (e.g.
Reipurth et al. 1999; Girart et al. 1997). (Note that this problem is
not confined to star formation: the direction of jets emanating from
the nuclei of Seyfert galaxies shows no correlation with the
orientation of the large scale galactic disk: e.g. Kinney et al. 2000;
Nagar \& Wilson 1999.)

Remarkably, for the most part, the outflows in the G35.2N core {\em
do} seem to have the same alignment; that is north-east to
south-west. The intriguing result in the current case is that it is
the most luminous source, G35.2N itself, which appears to be
discrepant as the axis of the G35MM2 flow is perpendicular to the
flattened core. Now the question to answer is why is the radio jet not
perpendicular to the long axis of the core, as would normally be
expected? Perhaps the formation of the other stars (particularly radio
source \#6), has exerted a sufficiently large torque to influence the
rotation axis of G35.2N. This would suggest that the other sources are
indeed high-mass YSOs. Theoretical work on such a scenario is not well
developed and so the above description remains rather speculative.

\section{Conclusions: a new picture of G35.2--0.7N}

In this paper we have presented a series of observations made with the
JCMT, the BIMA millimetre array and the VLA of the massive
star-forming region associated with G35.2--0.7N. The combination of
these data, along with a reinterpretation of previous observations has
led us to develop a new picture for the environment of G35.2N. While
superficially the CO 3\too 2 data support the hypothesis of a
jet-driven flow precessing through a large-angle, closer inspection
shows that there are probably as many as four outflows associated with
the sources in G35.2--0.7N. Much of the structure seen in the CO flow
can be explained as at least two overlapping flows. In addition, what
was previously believed to be a large-scale, rotating toroid around
G35.2N is reinterpreted here as being a rotating flattened, envelope
which is fragmented rather than continuous and contains a number of
YSOs embedded within dense clumps of gas and dust. VLA observations at
6- and 3.5 cm reveal several embedded YSOs, one of which lies within a
previously known ammonia clump.

The presumed most-massive YSO, G35.2N itself, is detected at 3.5 cm
and only weakly at 6 cm but appears to be distinguishable by virtue of
its positive spectral index and proximity to OH and water masers. The
radio jet is resolved into a number of individual knots, whose
distribution suggests that the jet is precessing with a half-angle of
approximately 10 degrees.  Surprisingly, some of the jet knots have
positive spectral indices while others have negative values. It is not
clear why they should differ so markedly. The northern lobe of the
outflow associated with the radio jet has been detected for the first
time in CO and possibly SiO. Dust emission from the envelope is
detected at 3.5 mm which has a dust-derived mass of 600 to 1400
M$_\odot$, depending on the assumed absorption coefficient. The dust
emission peaks 3 arcsec south-east of G35.2N, while the \htcop\ peaks
towards the location of a new millimetre source, G35MM2. G35MM2
marks the position of a deeply embedded massive YSO and is not
detected in our VLA images and is most likely the candidate for
driving much of the large-scale CO flow.

These results indicate the importance of high-resolution observations
in unravelling the processes involved in massive star formation. Even
now, higher-resolution dust and molecular line studies are necessary
to help further clarify the structure of G35.2--0.7N.

\vspace{5mm}
\noindent
{\bf ACKNOWLEDGMENTS:}

\noindent
AGG wishes to thank to Brian Murphy for assistance with the JCMT
observations and Greg Taylor at the VLA AOC, Socorro, for help
tracking down problems with the VLA data. John Carpenter is thanked
for providing data in advance of publication. The referee, Friedrich
Wyrowski, is thanked for his comments which helped clarify some points
in this paper. AGG acknowledges the financial support of PPARC for a
portion of this work. The James Clerk Maxwell Telescope is operated by
the Joint Astronomy Centre on behalf of the Particle Physics and
Astronomy Research Council of the United Kingdom, the Netherlands
Organization for Scientific Research and the National Research Council
of Canada. The National Radio Astronomy Observatory is operated by
Associated Universities, Inc., under cooperative agreement with the
National Science Foundation. BIMA is funded through grant AST-9981289
from the National Science Foundation.

\bsp


\begin{thebibliography}{}
\bibitem{} Allen C.W., 1973, Astrophysical Quantities. Athlone, London.
\bibitem{} Anglada G., Villuendas E., Estalella R., Beltr\'an M.T.,
Rodr\'{\i}guez L.F., Torrelles J.M., Curiel S., 1998, AJ, 116, 2953
\bibitem{} Arce H.G., Goodman A.A., 2001a, ApJ, 551, L171
\bibitem{} Arce H.G., Goodman A.A., 2001b, ApJ, 554, 132
\bibitem{} Bachiller R., Guilooteau S., Gueth F., Tafalla M., Dutrey
  A., Codella C., Castets A., 1998, A\&A, 339, L49
\bibitem{} Bally J., Lada C.J., 1983, ApJ, 265, 824
\bibitem{} Beuther H., Schilke P., Sridharan T.K., Menten K.M.,
  Walmsley C.M., Wyrowsk F., 2002, A\&A, 383, 892
\bibitem{} Bonnell I.A., Bate M.R., Clarke C.J., Pringle J.E., 1997,
MNRAS, 285, 201
\bibitem{} Brebner G.C., Cohen M.J., Heaton B.D., Davies S.R., 1987,
MNRAS, 229, 679
\bibitem{} Claussen M.J., Marvel K.B., Wootten A., Wilking B.A.,
1998, ApJ, 507. L79
\bibitem{} Cliffe J.A., Frank A., Jones T.W., 1996, MNRAS, 282, 1114
\bibitem{} Codella C., Bachiller R., Reipurth B., 1999, A\&A, 343, 585
\bibitem{} Davis C.J., Moriarty-Schieven G.M., Eisl\"offel J., Hoare
M.G., Ray T.P., 1998, AJ, 115, 1118
\bibitem{} Dent W.R.F., Little L.T., White G.J., 1984, MNRAS, 210, 173
\bibitem{} Dent W.R.F., Little L.T., Kaifu N., Ohishi M., Suzuki S.,
1985, A\&A, 146, 375
\bibitem{} Dent W.R.F., Little L.T., Sato S., Ohishi M., Yamashita T.,
  A\&A, 217, 217
\bibitem{} Dent W.R.F., Sandell G., Duncan W.D., Robson E.I., 1989,
MNRAS, 238, 1497 
\bibitem{} Downes T.P., Ray T.P., 1999, A\&A, 345, 977
\bibitem{} Eisl\"offel J., 1997, in Reipurth B., Bertout C., eds,
Herbig-Haro Flows and the Birth of Low Mass Stars. Kluwer, Dordrecht, p.93
\bibitem{} Fich M., Lada C.J., 1997, ApJ, 484, L63
\bibitem{} Forster J.R., Caswell J.L, 1999, A\&AS, 137, 43
\bibitem{} Fuller G.A., Zijlstra A.A., Williams S.J., 2001, ApJ, 555, L125
\bibitem{} Ghavamian P., Hartigan P., 1998, ApJ, 501, 687
\bibitem{} Gibb A.G., Davis C.J., 1998, MNRAS, 298, 644
\bibitem{} Gibb A.G., Heaton B.D., 1993, A\&A, 276, 511
\bibitem{} Girart J.M., Estalella R., Anglada G., Ho P.T.P.,
Rodr\'{\i}guez L.F., 1997, ApJ, 489, 734
\bibitem{} Goldreich P., Kwan J., 1974, ApJ, 189, 441
\bibitem{} Gueth F., Guilloteau S., 1999, A\&A, 343, 571
\bibitem{} Heaton B.D., Little L.T., 1988, A\&A, 195, 193
\bibitem{} Henriksen R., Ptuskin V.S., Mirabel I.F., 1991, A\&A, 248, 221
\bibitem{} Hoare M.G., 2002, in Crowther P.A., ed, The Earliest Phases
  of Massive Star Birth. San Francisco, The Astronomical Society of
  the Pacific, p.137 
\bibitem{} Hoare M.G., Drew J.E., Muxlow T.B., Davis R.J., 1994, ApJ,
421, L51
\bibitem{} Hollenbach D.J., McKee C.F., 1979, ApJS, 41, 555
\bibitem{} Hutawarakorn B., Cohen R.J., 1999, MNRAS, 303, 845
\bibitem{} Irvine W.M., Goldsmith P.F., Hjalmarson \AA, 1987, in
  Hollenbach D.J., Thronson H.A., Jr, eds, Interstellar
  Processes. Reidel, Dordrecht, p. 561
\bibitem{} Kinney A.L., Schmitt H.R., Clarke C.J., Pringle J.E.,
Ulvestad J.S., Antonucci R.R.J., 2000, ApJ, 537, 152
\bibitem{} Lada C.J., 1985, ARA\&A, 23, 267
\bibitem{} Lada C.J., Fich M., 1996, ApJ, 459, 638
\bibitem{} Lee C-F., Mundy L.G., Reipurth B., Ostriker E.C., Stone
J.M., 2000, ApJ, 542, 925
\bibitem{} Lee C-F., Stone J.M., Ostriker E.C., Mundy L.G., 2001, ApJ,
557, 429
\bibitem{} Little L.T., Dent W.R.F., Heaton B.D., Davies S.R., White
G.J., 1985, MNRAS, 217, 227
\bibitem{} Little L.T., Kelly M.L., Murphy B.T., 1998, MNRAS, 294, 105
\bibitem{} Mart\'{\i} J., Rodr\'{\i}guez L.F., Reipurth B., 1993, ApJ,
416, 208
\bibitem{} Masson C.R., Chernin L.M., 1992, ApJ, 387, L47
\bibitem{} Matzner C.D., McKee C.F., 1999, ApJ, 526, L109
\bibitem{} Mezger P.G., Chini R., Kreysa E., Wink J., 1987, A\&A, 182, 127
\bibitem{} Nagar N.M., Wilson A.S., 1999, ApJ, 516, 97
\bibitem{} Panagia N., Felli M., 1975, A\&A, 39, 1
\bibitem{} Reipurth B., Yu K.C., Rodr\'{\i}guez L.F., Heathcote S.,
Bally J., 1999, A\&A, 352, L83
\bibitem{} Richer J.S., Hills R.E., Padman R., 1992, MNRAS, 254, 525
\bibitem{} Richer J.S., Shepherd D.S., Cabrit S., Bachiller R.,
  Churchwell E., 2000, in Mannings V., Boss A.P., Russell S.S., eds,
  Protostars \& Planets IV. University of Arizona Press, Tucson, p.867
\bibitem{} Rodr\'{\i}guez L.F., 1997, in Reipurth B., Bertout C., eds,
Herbig-Haro Flows and the Birth of Low Mass Stars. Kluwer, Dordrecht, p.83
\bibitem{} Schilke P., Walmsley C.M., Pineau des For\^ets G., Flower
D.R., 1997, A\&A, 321, 293
\bibitem{} Shepherd D.S., Watson A.M., Sargent A.I., Churchwell E.,
1998, ApJ, 507, 861
\bibitem{} Shepherd D.S., Yu K.C., Bally J., Testi L., 2000, ApJ, 535, 833
\bibitem{} Simon M., Felli M., Massi M., Cassar L., Fischer J., 1983,
  ApJ, 266, 623
\bibitem{} Torrelles J.M., G\'{o}mex J.F., Rodr\'{\i}guez L.F., Curiel
  S., Ho P.T.P., Garay G., 1996, ApJ, 457, L107
\bibitem{} Vall\'ee J.P., Bastien P., 2000, ApJ, 530, 806
\bibitem{} Walther D.M., Aspin C.A., McLean I.S., 1990, ApJ, 356, 544
\bibitem{} Wilner D.J., Reid M.J., Menten K.M., 1999, ApJ, 513, 775
\bibitem{} Wright A.E., Barlow M.J., 1974, MNRAS, 170, 41
\end{thebibliography}
\end{document}